\documentclass[twocolumn,prd,aps,tightenlines,floats,floatfix,preprintnumbers,nofootinbib,eqsecnum]{revtex4-2}%%

\def\sp{\kern +3pt}
\def\sm{\kern -7pt}
\def\spQ{\kern +6pt}

\def\bea{\begin{eqnarray}}
\def\eea{\end{eqnarray}}

\def\sfrac#1#2{{\textstyle \frac{#1}{#2}}}

\def\be{\begin{equation}}
\def\ee{\end{equation}}
\def\ba{\begin{eqnarray}}
\def\ea{\end{eqnarray}}

\usepackage{graphics}
\usepackage{graphicx}
\usepackage{epsf}
\usepackage{amsmath}
\usepackage{amssymb}
\usepackage{bbold}
\usepackage{xcolor}

\usepackage{fnbreak}

\begin{document}

\phantom{0}
\vspace{-0.2in}
\hspace{5.5in}

% include preprint number option
%\preprint{{\bf  }}

\vspace{-1in}%\parbox{1.5in}{ \vspace{-9.6in}}  % moves the preprint box down

\title
{\bf About the magnitude of the $\gamma^\ast N \to N(1520)$
transverse \\ amplitudes near $Q^2=0$}
\author{G.~Ramalho}
\vspace{-0.1in}

\affiliation{Department of Physics and OMEG Institute, Soongsil University, \\
Seoul 06978, Republic of Korea}

\vspace{0.2in}
\date{\today}

\phantom{0}

\begin{abstract}
The  $\gamma^\ast N \to N(1520)$ transition has a property
that differs from the other low-lying
nucleon resonance amplitudes:
the magnitude of the transverse helicity amplitudes.
The transition helicity amplitudes are defined
in terms of square-transfer momentum $q^2$, or $Q^2=-q^2$.
Near the photon point ($Q^2=0$)
there is a significant difference in the magnitude
of the transverse amplitudes: $A_{3/2}$ 
is very large and $A_{1/2}$ is very small.
This atypical behavior contrasts with the
relation between the amplitudes at the
pseudothreshold [the limit where the nucleon
and the $N(1520)$ are both at rest and $Q^2 <0$],
where $A_{3/2} = A_{1/2}/\sqrt{3}$, and also in 
the large-$Q^2$ region, where theory and data suggest
that $A_{3/2}$ is suppressed relative to $A_{1/2}$.
In the present work, we look for the source
of the suppression of the $A_{1/2}$ amplitude at $Q^2=0$.
The result is easy to understand in first approximation,
when we look into the relation between the transverse amplitudes
and the elementary form factors, defined by a gauge-invariant parametrization
of the $\gamma^\ast N \to N(1520)$ transition current, near $Q^2=0$.
There is a partial cancellation between contributions of 
two elementary form factors near $Q^2=0$.
We conclude, however, that the correlation between
the two elementary form factors at $Q^2=0$ 
is not sufficient to explain the transverse amplitude
data below $Q^2 = 1$ GeV$^2$.
The description of the dependence of the transverse amplitudes on $Q^2$ 
requires the determination of the scale of variation
of the elementary form factors in the range $Q^2=0$...0.5 GeV$^2$,
a region with almost non existent data.
We conclude at the end that the low-$Q^2$ data for the transverse amplitudes
can be well described when we relate the scale of variation 
of the elementary form factors with
the nucleon dipole form factor.
\end{abstract}

%\phantom{0}
%\vspace{7.0in}
%\vspace{-6in}
\vspace*{0.9in}  % sets how far the title is below the preprint box
\maketitle

\section{Introduction}
\label{secIntro}

In the last two decades there was a significant progress
in the experimental study of the electromagnetic
structure of the nucleon ($N$) and the nucleon resonances ($N^\ast$).
The helicity amplitudes associated with the 
$\gamma^\ast N \to N^\ast$ transitions have been measured 
in detail for the $\Delta(1232)$,
$N(1440)$, $N(1520)$,
and $N(1535)$ resonances in a range from $Q^2=0.25$ GeV$^2$ up to
4 or 6 GeV$^2$~\cite{PPNP2023,NSTAR,Aznauryan12,CLAS23a,Drechsel07a,Burkert04}.
The measured helicity amplitudes are: 
the transverse amplitudes $A_{1/2}$ and $A_{3/2}$
(for spin $J \ge 3/2$) and the
longitudinal amplitude $S_{1/2}$. 
Near the photon point, however, there are still
some uncertainties associated with the shape associated with 
the helicity amplitudes~\cite{PPNP2023,Siegert4,Tiator16a}.
The selection from the Particle Data Group (PDG) at $Q^2=0$
has a large band of variation~\cite{PDG2022},
and for most resonances there are no data below
$Q^2=0.25$ GeV$^2$~\cite{PPNP2023,Mokeev22a}.

Among the best known experimental resonances
the  $N(1520)\frac{3}{2}^-$
(spin $J=\frac{3}{2}$ and negative parity, $P=-$)
has properties that differ
from the other low-lying nucleon excitations.
The transverse amplitudes $A_{1/2}$ and $A_{3/2}$ have completely
different magnitudes near the photon point~\cite{PDG2022}, 
and the helicity amplitudes are
related by two conditions near the pseudothreshold point, 
where $Q^2=-(M_R-M)^2$~\cite{PPNP2023,Devenish76}
($M$ is the mass of the nucleon and $M_R$
is the mass of the nucleon resonance).
Most transitions are constrained by only
one condition~\cite{PPNP2023,Siegert4,Devenish76}.
Although these constraints are valid in a region
not directly accessed by electron scattering on nucleons ($Q^2 < 0$),
that may not be probed directly in physical experiments,
the relations may have a significant impact
on the shape of the helicity amplitudes at low $Q^2$, when
the masses of the nucleon and the nucleon resonance are 
close~\cite{PPNP2023,Siegert4,Tiator16a,Siegert3,Siegert2,Siegert5}.
Numerically, the pseudothreshold occurs when $Q^2 \simeq - 0.38$ GeV$^2$.

In the present work, we study the magnitude of the
$\gamma^\ast N \to N(1520)$ transverse amplitudes near $Q^2=0$,
based on the analytic structure of the transition current
and on the correlations between the amplitudes in the low-$Q^2$ region.
We start by reviewing what we know about
the transverse amplitudes in three kinematic regions.

Near the pseudothreshold, in addition to
the condition associated with Siegert's
theorem~\cite{Siegert-note,Tiator16a,Drechsel92a,Buchmann98a},
one has the relation
$A_{3/2} = \sqrt{3} A_{1/2}$~\cite{PPNP2023,Devenish76,Siegert4}.
In the large-$Q^2$ region, theoretical calculations 
based on constituent quark-counting rules and perturbative QCD arguments
indicate that there is a strong dominance of the
$A_{1/2}$ amplitude over the $A_{3/2}$ amplitude
($|A_{1/2}| \gg |A_{3/2}|$)~\cite{PPNP2023,Aznauryan12,Drechsel07a,Warns90a}.
Finally, near $Q^2=0$, one can quote the information from 
the PDG~\cite{PDG2022} 
\ba
& &A_{3/2} = +(140 \pm 5)\times 10^{-3} \; \mbox{GeV}^{-1/2}, \nonumber \\
& &A_{1/2} = -(22.5 \pm 7.5)\times 10^{-3} \; \mbox{GeV}^{-1/2}.
\label{eqAmpsEXP}
\ea
From these results, we can conclude that there is a considerable suppression of $A_ {1/2} $ relative to $A_ {3/2} $ at the photon point.

We can summarize our knowledge of the ratio  $A_{1/2}/A_{3/2}$,
in the three regimes, as
\ba
{\cal R}  \equiv 
\frac{A_{1/2}}{A_{3/2}}
=
\left\{
\begin{array}{ccc}
  \frac{1}{\sqrt{3}} & \mbox{if} & Q^2= -(M_R-M)^2 \cr
        \vspace{.1cm}
               - \epsilon & \mbox{if} & Q^2=0 \cr
               \vspace{.1cm}
       \infty & \mbox{if} & Q^2=+ \infty \cr
\end{array},
\right.
\ea
where $\epsilon$ represents a small positive value,
$\epsilon \simeq 0.18 \simeq \frac{1}{5}$, 
according to the experimental data (\ref{eqAmpsEXP}).

At the pseudothreshold, the amplitudes have similar magnitudes
(${\cal R} \simeq 0.6$).
The suppression of $A_{3/2}$ at large-$Q^2$
is extensively discussed
in the literature~\cite{PPNP2023,Aznauryan12,Warns90a,N1520}.
The theoretical challenge is then to understand
why is the ratio between the two amplitudes
so small in absolute value near $Q^2=0$.

From the theoretical point of view, there is some debate
about the nature of the $N(1520)$ resonance:
if it is dominated by valence quark degrees of freedom,
or alternatively, if it is dominated by baryon-meson
molecular-like states~\cite{PPNP2023,N1520,Burkert04,CLAS16}.
The magnitude of $A_{3/2}(0)$ is difficult to explain based solely
on the quark core structure of the baryon states.
Quark model calculations explain in general
only about one-third or one-half of
the measured value of the amplitude~\cite{Warns90a,Merten02,Ronniger13a,Golli13a,Giannini15a}. 
Those estimates are improved when explicit meson cloud dressing
or quark-antiquarks excitations are taken into account
in quark model calculations~\cite{Golli13a,Bijker09a}.
Calculations based on dynamical coupled-channel reaction models,
where the baryon resonances are described in terms 
of baryon-meson states~\cite{Burkert04,JDiaz08,Kamano13},
predict large contributions to the amplitude $A_{3/2}$ at low $Q^2$,  
on the order of 50\% of the experimental values~\cite{CLAS16}.
In the present work, we look for the origin of
the difference of magnitudes between $A_{1/2}$ and $A_{3/2}$,
based on the numerical contributions for each amplitude, 
without an explicit reference to the internal degrees of freedom.

The transverse amplitudes can be expressed in
terms of the multipole form factors: the magnetic dipole ($G_M$) and 
the electric quadrupole ($G_E$) form factors,
as defined by Devenish {\it et al.}~\cite{PPNP2023,Aznauryan12,Devenish76},
\ba
A_{1/2}= - \frac{1}{4 F}  T_1,  \hspace{0.9cm} 
A_{3/2}= - \frac{\sqrt{3}}{4 F} T_2,
\label{eqAmps-v1}
\ea
where
\ba
T_1 \equiv G_E - 3 G_M,  \hspace{0.9cm}
T_2 \equiv G_E + G_M,
\label{eqAmps-v2}
\ea
and the factor $F$ takes the form
\ba
F = \frac{1}{e} \frac{2M}{M_R -M} \sqrt{\frac{M M_R K}{(M_R +M)^2 + Q^2}},
\label{eqFm}
\ea
with $K= \frac{M_R^2-M^2}{2 M_R}$,  $e = \sqrt{4 \pi \alpha}$, and 
$\alpha \simeq 1/137$ is the hyperfine structure constant.

From the previous relations, we can conclude that $A_{1/2} \simeq 0$, 
near $Q^2=0$
is equivalent to the result $G_E \simeq 3 G_M$.
Notice, however, that this analysis 
only transfers the discussion from helicity amplitudes
to the multipole form factors $G_E$ and $G_M$,
and tells us nothing about the correlation between $G_E$ and $G_M$.

The results $A_{1/2} \simeq 0$ or  $G_E \simeq 3 G_M$
can be understood when we write the relations between 
the helicity amplitudes and the multipole form factors
in terms the elementary form factors,
defined by the gauge-invariant representation 
of the transition current for a $J^P= \frac{3}{2}^-$ nucleon resonance.
The transition current can 
be expressed in terms of three independent gauge-invariant structures
which define three independent forms factors
that can be labeled as $G_1$, $G_2$, and $G_3$,
and are free of kinematic singularities~\cite{PPNP2023,Devenish76}.
For convenience, we call these functions elementary form factors.

Using the elementary form factors,
we can re-write the transverse amplitudes (\ref{eqAmps-v1})
in the limit $Q^2=0$, as
\ba
A_{1/2} &=&  - \frac{1}{4 F_0} \, T_1,   \label{eqA12b}\\
A_{3/2} &=&  - \frac{\sqrt{3}}{4 F_0} \,
\left[T_1 - 4 \frac{M}{\sqrt{6}}\frac{M_R-M}{M_R} G_1 \right],
\label{eqA32b}
\ea
where $F_0 = {\cal B} \sqrt{M}$
and ${\cal B} = \frac{1}{e} \frac{M}{M_R} \sqrt{\frac{M_R}{K}} \simeq 3.67$
is dimensionless. 
The factor $T_1$, defined by Eqs.~(\ref{eqAmps-v2}), takes the form
\ba
T_1 = - 4\frac{M}{\sqrt{6}}
  \left[\frac{M}{M_R} G_1 + \frac{1}{2} (M_R + M) G_2 \right].
\label{eqT1a}
\ea

From the relations (\ref{eqA12b})--(\ref{eqT1a}), 
we can then conclude that in the limit $Q^2=0$,
the transverse amplitudes depend only on the
values of the functions $G_1$ and $G_2$.
We can also conclude that $A_{1/2} \simeq 0$,
when $T_1$ is negligible, and as a consequence
$A_{3/2} \propto \frac{M_R-M}{M_R} G_1$ is large when $G_1$ is large.
The numerical result for $A_{1/2}$ is then explained when
$G_1$ and $G_2$ are large and have opposite signs.
In this case, there is a significant cancellation
between the terms in $G_1$ and in $G_2$.
We will conclude, however, that $T_1 \simeq 0$ ($\epsilon \simeq 0$) provide
only a rough explanation of the data.
The values of $G_1$ and $G_2$ at $Q^2=0$ have
corrections of the order of 30\% and 20\%, respectively, 
when we use the experimental ratio $A_{1/2}/A_{3/2}$ ($\epsilon \simeq 0.2$),
instead of $A_{1/2}=0$ ($\epsilon =0$).

At this point, one can ask if the values
of $G_1(0)$ and $G_2(0)$ can help to explain 
the $Q^2$ dependence of $A_{1/2}$ and $A_{3/2}$
in the range $Q^2=0$...1 GeV$^2$.
A simple numerical calculation demonstrates, however,
that the shape of the amplitude $A_{3/2}$ cannot be
explained without an estimate of the derivative
of the elementary form factors $G_i$.
We conclude at the end that the $A_{1/2}$ and $A_{3/2}$ data
can be well described when we consider simple multipole
parametrizations of the form factors $G_i$,
where the scale of variation is determined by the
scale of the nucleon dipole form factor,
used in parametrizations of the nucleon electromagnetic
form factors and some $\gamma^\ast N \to N^\ast$ transition form factors.

We propose parametrizations of the $A_{1/2}$ and $A_{3/2}$
amplitudes based on our analysis of the
amplitudes at $Q^2=0$.
The parametrizations are consistent with the $Q^2=0$...1 GeV$^2$ data, 
within the uncertainties of the available data,
and may be tested by future  experiments 
in facilities like MAMI or JLab-12 GeV in the low-$Q^2$ region~\cite{Mokeev22a}.
The precision of the present estimates can be improved once
the uncertainties of the $A_{1/2} (0)$ and $A_{3/2} (0)$ data are reduced.

This article is organized as follows:
in the next section we present the general formalism for the 
$\gamma^\ast N \to N^\ast$ transition form factors and helicity amplitudes
for $J^P= \frac{3}{2}^-$ nucleon resonances,
and discuss the relevant limits (pseudothreshold, photon point and large $Q^2$).
Our numerical analysis of the elementary form factors at the photon point
is presented in Sec.~\ref{secResults1}.
In Sec.~\ref{secResults2}, we derive parametrizations
of the data based on the our analysis and discuss
the limits of the parametrizations.
We finalize in Sec.~\ref{secResults2} with the outlook and conclusions.

\section{Helicity amplitudes and transition form factors}
\label{secFormalism}

We discuss now the formalism associated with
the $\gamma^\ast N \to N(1520)$ transition, and
the definition of helicity amplitudes and multipole form factors.

Considering an initial nucleon with the momentum $p$
and a final nucleon resonance with momentum $p'$,
we can define
\ba
q= p' -p, \hspace{.6cm} P= \sfrac{1}{2}(p'+ p),
\ea
as the transfer momentum and the average of the baryons momentum,
respectively.

The transition current between a nucleon and an $N^\ast$
$J^P = \frac{3}{2}^-$ state can be written as 
\ba
J^\mu= \bar u_\alpha (p') \Gamma^{\alpha \mu} (P,q)  \gamma_5 u(p),
\label{eqJ1}
\ea
where $u_\alpha$, $u$ are the resonance, and the nucleon spinors,
respectively, 
and $\Gamma^{\alpha \mu}$
takes the form~\cite{PPNP2023,Aznauryan12,Devenish76,N1520,SemiRel}
\ba
\Gamma^{\alpha \mu} (P,q) &=&  \left( q^\alpha \gamma^\mu - {\not \! q} g^{\alpha \mu} \right)
G_1 + \nonumber \\
& & \left[ q^\alpha P^\mu - (P \cdot q) g^{\alpha \mu} \right]  G_2 + \nonumber \\
& &  \left( q^\alpha q^\mu - q^2 g^{\alpha \mu} \right) G_3.
\label{eqJ2}
\ea
In the previous relation, $G_i$ ($i=1,2,3$) are independent
functions, free of kinematic singularities, 
refereed to hereafter as elementary form factors.
Comparatively with other authors that use
the Devenish convention for the operators,
and define the second term of Eq.~(\ref{eqJ2})
in terms of $p'= P + \sfrac{1}{2} q$~\cite{PPNP2023,Aznauryan12,Devenish76},
we follow the Jones and Scadron convention~\cite{Jones73}
and use $P$ to define the operator associated with $G_2$~\cite{Jones73,N1520,SemiRel,Siegert3}.
The conversion is trivial\footnote{To obtain the Devenish
form factors~\cite{Devenish76}
in terms of the Jones and Scadron form factors~\cite{Jones73}
we replace
$G_1 \to G_1$, $G_2 \to G_2$, and $G_3 \to G_3 + \sfrac{1}{2}G_2$.}.

For the representation of the helicity amplitudes,
defined at the resonance rest frame,
it is convenient to introduce the magnitude of the transfer three-momentum $|{\bf q}|$.
This variable can be written in a covariant form as 
\ba
|{\bf q}| = \frac{\sqrt{Q_+^2 Q_-^2}}{2 M_R},
\ea
using the notation
\ba
Q_\pm^2= (M_R \pm M)^2 + Q^2.
\ea

The magnetic dipole ($G_M$) and the electric quadrupole ($G_E$)
form factors can be
calculated inverting the relations (\ref{eqAmps-v1}) and
(\ref{eqAmps-v2})
\ba
G_M & = & - F \left[\sfrac{1}{\sqrt{3}} A_{3/2} - A_{1/2} \right],
\label{eqGM}\\
G_E  & =&  - F \left[\sqrt{3}  A_{3/2} + A_{1/2} \right].
\label{eqGE}
\ea

One can also relate the longitudinal (scalar) amplitude $S_{1/2}$
with the Coulomb quadrupole form factor $G_C$, 
\ba
S_{1/2} = - \frac{1}{\sqrt{2} F} \frac{|{\bf q}|}{2 M_R} G_C.
\label{eqS12}
\ea

Using the expressions (\ref{eqJ1}) and (\ref{eqJ2}),
we can write the magnetic dipole and the electric quadrupole form factors
in terms of $G_i$, as~\cite{PPNP2023,Aznauryan12}
\ba
G_M &= &- Z_R \, Q_-^2 \frac{G_1}{M_R}, \label{eqGE1} \\
G_E &= & -Z_R  \left\{
    [ (3M_R + M) (M_R -M) -Q^2] \frac{G_1}{M_R} \right.   \nonumber \\
& & \left. \frac{}{} + 2 (M_R^2 - M^2)G_2 - 4 Q^2 G_3\right\} ,
\label{eqGM1}
\ea
where
$Z_R= \frac{1}{\sqrt{6}} \frac{M}{M_R -M}$.

Using the previous equations,
we conclude that
\ba
T_1 &= & - Z_R \left\{
4 [M(M_R-M) - Q^2] \frac{G_1}{M_R}  \right.
\nonumber \\
& &  \left. \frac{}{} + 2 (M_R^2 - M^2) G_2 - 4 Q^2 G_3 \right\},
\label{eqT1x}\\
T_2 & =& - Z_R \left\{ 4 (M_R-M) G_1  \right.  \nonumber \\
& & \left. +   2 (M_R^2 - M^2) G_2 - 4 Q^2 G_3 \right\}.
\label{eqT2x}
\ea

We can also write
\ba
T_2 &=& T_1 - 4 Z_R Q_-^2 \frac{G_1}{M_R} \nonumber \\
& = & T_1 + 4 G_M.
\label{eqT1T2a}
\ea

For future discussion, we write also the
relation between the Coulomb quadrupole form factor
and the elementary form factors, 
\ba
G_C &= & Z_R \left[ 4 M_R G_1 + (3 M_R^2 + M^2 + Q^2) G_2
  \right. \nonumber \\
  & & \left. + 2(M_R^2- M^2-Q^2) G_3 \right] .
\label{eqGC}
\ea
The previous relation can be used to calculate the amplitude $S_{1/2}$, 
according to Eq.~(\ref{eqS12}).
Notice that $S_{1/2}$ and $G_C$ cannot be measured
at the photon point (because there are no real photons with zero polarization).
The relation (\ref{eqGC}) can be used, however,
to estimate $G_C$ and $G_3$ for values
of $Q^2$ arbitrarily close to $Q^2=0$.

We discuss now briefly the three relevant limits:
the pseudothreshold, the photon point and the large-$Q^2$ limit.

\subsection{Pseudothreshold \label{secFormalismA}}

As mentioned already,
when we study the electromagnetic properties
based on the helicity amplitudes or the multipole form factors,
there are some conditions between those functions that need to be fulfilled
when we consider the pseudothreshold
limit \mbox{$Q^2= -(M_R-M)^2$}~\cite{PPNP2023,Devenish76}.
These conditions are the consequence of the
gauge-invariance structure of the transition current,
which requires that the elementary form factors
are independent and free of kinematic
singularities~\cite{Devenish76,Jones73}.

There are two conditions to be considered for the
$\gamma^\ast N \to N\left(\sfrac{3}{2}^- \right)$ 
multipole transition form factors~\cite{Devenish76,Siegert3}:
\ba
G_M \propto |{\bf q}|^2, \hspace{.7cm}
G_C = - \frac{M_R -M}{M_R} G_E.
\label{eqSiegertFF}
\ea
These conditions can be transposed to the helicity amplitudes,
as~\cite{Siegert4,Tiator16a}
\ba
& &
A_{3/2} = \sqrt{3} A_{1/2},
\label{eqSiegertAmp1}\\
& &
(A_{1/2} + \sqrt{3} A_{3/2}) = - 2 \sqrt{2} (M_R -M) \frac{S_{1/2}}{|{\bf q}|}.
\label{eqSiegertAmp2}
\ea
In addition, it is expected that
$S_{1/2} \propto {\cal O}(|{\bf q}|)$ and 
$A_{1/2}$, $A_{3/2} \propto {\cal O}(1)$, 
near $|{\bf q}|=0$~\cite{Siegert4,Drechsel92a}.

The correlation between the transverse amplitudes (\ref{eqSiegertAmp1})
is equivalent to the relation $G_M=0$ from (\ref{eqSiegertFF})
when $|{\bf q}|=0$.

The second condition for the helicity amplitudes
relates the electric amplitude,
$E \equiv (A_{3/2} + \sqrt{3} A_{1/2})$, with
the scalar amplitude $S_{1/2}$, and correspond
to Siegert's theorem for the $J^P= \frac{3}{2}^-$ nucleon 
resonances~\cite{PPNP2023,Siegert1,DeForest66a,Amaldi79a,Drechsel92a,Buchmann98a,Tiator16a,Siegert5}.

Using the relations between the helicity amplitudes and the multipole form factors
(\ref{eqGE1}) and (\ref{eqGM1}), and $G_M =0$,
we can conclude that
\ba
A_{3/2} = \sqrt{3} A_{1/2} = - \frac{\sqrt{3}}{4 F} G_E,
\ea
where $F= \frac{1}{e} \frac{M}{\sqrt{2 M_R}} \sqrt{\frac{M_R + M}{M_R -M}}$.

The conditions (\ref{eqSiegertFF}) for the form factors 
are valid for the nucleon resonances
$J^P= \frac{3}{2}^-, \frac{5}{2}^+, \frac{7}{2}^-, ...$.
Modified versions of the conditions for the helicity
amplitudes (\ref{eqSiegertAmp1}) and  (\ref{eqSiegertAmp2}) 
are also valid for $J^P= \frac{5}{2}^+, \frac{7}{2}^-, ...$~\cite{PPNP2023}.
Among all those nucleon resonances,
the  $N(1520)\frac{3}{2}^-$ resonance is one
of the resonances with stronger impact of the pseudothreshold conditions
on parametrizations compatible
with the available data~\cite{Siegert4},
due to the proximity between pseudothreshold and photon points.

\subsection{Photon point}

In the limit $Q^2=0$, we can write
\ba
\hspace{-1.3cm}
T_1 &=& -  4 \frac{M}{\sqrt{6}} \left\{ \frac{M}{M_R} G_1 + \frac{1}{2} (M_R+  M) G_2 \right\}, \label{eqT1b}\\
\hspace{-1.3cm}
T_2 &=& -4 \frac{M}{\sqrt{6}}  \left\{G_1 + \frac{1}{2} (M_R + M) G_2 \right\} .
 \label{eqT2b}
\ea
We can also write, following Eq.~(\ref{eqT1T2a})
\ba
T_2 &=& T_1 - 4 \frac{M}{\sqrt{6}} (M_R -M)\frac{G_1}{M_R} \nonumber \\
& = & T_1 + 4 G_M,
\label{eqT1T2b}
\ea
and
\ba
G_M = - \frac{M}{\sqrt{6}} (M_R -M) \frac{G_1}{M_R}.
\label{eqT2c}
\ea
To obtain the previous relations, we used
$(M_R -M) Z_R= \frac{M}{\sqrt{6}}$.

Concerning the scalar amplitude, we can write
\ba
& &F_0 S_{1/2} (0) =
- \frac{1}{2 \sqrt{3}} \frac{M(M_R + M)}{4 M_R^2} \nonumber \\
& &\times [ 4 M_R G_1 + (3 M_R^2+ M^2)G_2 + 2 (M_R^2-M^2) G_3 ]. \nonumber \\
\ea

\subsection{Large $Q^2$}

The large-$Q^2$ region has been discussed in detail in 
the literature~\cite{PPNP2023,Aznauryan12}.
Here, we present the summary.
At large $Q^2$ the transverse amplitudes follow~\cite{PPNP2023,Carlson8X,Carlson98b}
\ba
A_{1/2} \propto \frac{1}{Q^3}, \hspace{.6cm}
A_{3/2} \propto \frac{1}{Q^5},
\label{eqAmpsLQ2}
\ea
meaning that $A_{3/2}$ is suppressed relatively
to $A_{1/2}$.

The corresponding relations for the form factors are~\cite{PPNP2023,N1520}
\ba
& & G_E \propto \frac{1}{Q^4}, \hspace{.3cm} G_M \propto \frac{1}{Q^4}, \label{eqFFLQ2a}\\
& & G_E = - G_M + {\cal O} \left( \frac{1}{Q^6}\right).
\label{eqFFLQ2b}
\ea

\section{Form factors $G_i$ for $Q^2=0$
 \label{secResults1}}

In the analysis of the transverse amplitudes
near \mbox{$Q^2=0$}, we consider different approximations.
For the discussion, we convert the experimental data (\ref{eqAmpsEXP}),
into the dimensionless variables
\ba
& &
\tilde A_{3/2}= F_0 A_{3/2}(0) = + 0.498 \pm 0.018,
\label{eqAmps-scalar1} \\
& &\tilde A_{1/2}= F_0 A_{1/2}(0) = -0.080 \pm 0.027,
\label{eqAmps-scalar2}                
\ea
based on the numerical result $F_0 = 3.67 \sqrt{M}$.

\begin{table*}[th]
\begin{tabular}{l |  c c c |  c c c | cc | c}
\hline
\hline
&  $M G_1 (0)$ &  $M^2 G_2(0)$ & $M^2 G_3(0)$  & $A_{3/2}(0)$  & $A_{1/2}(0)$ & $S_{1/2}(0)$ & $G_M(0)$
& $G_E(0)$ & Label\\
\hline
  $T_1=0$      &   $2.507$    &   $-1.429$    &             & 140  & 0.0   &   & $-0.287$ & $-0.862$ \\[.2cm]
%\hline
%    & & &  & & &&  &\\
$T_1 = 0.319$   &   $2.354$     &   $-1.260$       &    $\; \; \; 0.000$   &  140 & $-22.5$  &  $-82.7$  &
$-0.367$ &  $-0.782$ &
Multipole 2a\\
                &   $2.354$     &   $-1.260$       &   $-0.140$   &  140 & $-22.5$  &  $-73.5$  &  $-0.367$ &  $-0.782$ & Multipole 2b \\
                &   $2.354$     &   $-1.260$       &   $-0.278$   &  140 & $-22.5$  &  $-64.4$  &    $-0.367$ &  $-0.782$ &  Multipole 2c\\
                 &   $(0.183)$      & \sp \sp$(0.134)$   &      &          &          &  &  & &{}\\
\hline
\hline
\end{tabular}
\caption{\footnotesize Model parameters $G_1(0)$, $G_2(0)$,
  according with the 
  values for the amplitudes $A_{3/2}(0)$, $A_{1/2}(0)$.
  We include also the limit of $S_{1/2} (Q^2)$ for $Q^2=0$,
  based on the values of $G_3 (0)$.
  The amplitudes are in units $10^{-3}$ GeV$^{-1/2}$.
  In the last row, the values between commas are the uncertainties
  of  $M G_1(0)$ and $M^2 G_2(0)$, based on the data for the transverse amplitudes.
  The model with $T_1 \ne 0$ has $T_2= -1.149$.
 }
\label{table-Models}
\end{table*}

Notice that, since the comparison between
amplitudes is made in units $10^{-3}$ GeV$^{-1/2}$,
the first quantity  
($\simeq 500 \times 10^{-3}$) can be regarded as a large number,
and the second quantity ($\simeq  -80 \times 10^{-3}$) can be regarded 
as a small number.

We can use the results (\ref{eqAmps-scalar1}) and
(\ref{eqAmps-scalar2}), to calculate the corresponding
form factors $G_1$ and $G_2$ for $Q^2=0$.
Inverting the relations (\ref{eqA12b})--(\ref{eqT1a}),
one obtains
\ba
& &
M G_1 = {\cal R} \left( \frac{1}{\sqrt{3}} \tilde A_{3/2} - \tilde A_{1/2}
\right), \label{eqG1a}\\
& &
M^2 G_2 = - \frac{2 M}{M_R+ M} \frac{M}{M_R} {\cal R}
\left( \frac{1}{\sqrt{3}} \tilde A_{3/2} - \frac{M_R}{M} \tilde A_{1/2}
\right), \nonumber \\
\label{eqG2a}
\ea
where
${\cal R} = \frac{\sqrt{6} M_R}{M_R -M}$.
The factors $M$ and $M^2$ are included to
generate dimensionless expressions.

In the introduction, we discussed the approximation 
$\tilde A_{1/2} =0$ ($T_1=0$), based on Eqs.~(\ref{eqA12b}) and (\ref{eqA32b}).
In that case, we obtain $G_2 = - \frac{2}{M_R + M}\frac{M}{M_R} G_1$.
Now, we can notice, using the relations (\ref{eqG1a}) and (\ref{eqG2a}), 
that the condition $T_1=0$ provides only a rough approximation,
since  $\tilde A_{1/2}$ is combined in fact with $\tilde A_{3/2}/\sqrt{3}$.
Neglecting $\tilde A_{1/2}$
in the estimates of $G_1$ and $G_2$ has an
impact of 28\% for $G_1$ and of 17\% for $G_2$.

The relations (\ref{eqG1a}) and (\ref{eqG2a})
can also be used to explain the significant cancellation in $T_1$.
The effect can be observed when we write $T_1$ on the form 
$T_1= - 4\frac{M}{\sqrt{6}} t_1$,
where
\ba
t_1 = \frac{M}{M_R} G_1 + \frac{1}{2} (M_R + M) G_2.
\ea
For that purpose, we write the two terms as 
\ba
\frac{M}{M_R} G_1 &=& \frac{{\cal R}}{M_R}
\left( \frac{1}{\sqrt{3}} \tilde A_{3/2} - \tilde A_{1/2}
\right), \nonumber \\
\frac{1}{2} (M_R + M) G_2 &= & - \frac{{\cal R}}{M_R}
\left( \frac{1}{\sqrt{3}} \tilde A_{3/2} - \tilde A_{1/2}
\right) \nonumber \\
& & + \frac{{\cal R}}{M_R} \frac{M_R-M}{M} \tilde A_{1/2}.
\nonumber 
\ea
In this form, one concludes that the first term of
\mbox{$\frac{1}{2} (M_R + M) G_2$} cancels the term in $G_1$, 
and only the term proportional to $\tilde A_{1/2}$
survives the sum.
The correction term is 13\% of the term in $G_1$.
In units $10^{-3}$ the term in $G_1$ and
the term in $G_2$ are large numbers with opposite signs.

The dominance of the amplitude $A_{3/2}$ is still explained
by the small magnitude of $T_1$.
When we can neglect $T_1$ in Eq.~(\ref{eqT1T2b}),
we conclude that $T_2 \propto G_1$.
Thus, the amplitude $A_{3/2}$ is large when $T_2$ is large, and $T_1$ is small
in comparison with $T_2$.
However, when we look for (\ref{eqT2b}): $T_2 = - 4\frac{M}{\sqrt{6}} t_2$,
with $t_2 = G_1 + \frac{1}{2} (M_R + M) G_2$, 
we conclude that $T_2$ is large because there
is only a partial cancellation between the two large terms.

To summarize, the combination of the results for
the transverse amplitudes is a consequence
of the large magnitude of the form factors $G_1$ and $-G_2$.
In $A_{1/2}$, one has a significant cancellation between
the term in $G_1$ and the term in $G_2$.
In $A_{3/2}$, the term in $G_1$ is enhanced and
the suppression between the terms is attenuated.
We conclude also that in first approximation
(leading order in $\tilde A_{1/2}$), one has $A_{3/2} \propto G_1$.

The values of $G_1$ and $G_2$ for $Q^2=0$ are
presented in Table~\ref{table-Models} for
the cases $T_1=0$ and $T_1 \ne 0$. 
The first row ($T_1 = 0$) gives the results
when we use  $A_{1/2}(0)=0$, 
and the second row gives the result
when $T_1 \ne 0$ is fixed by the experimental value of $A_{1/2}(0)$.
In the second row, we include also $G_3(0)=0$.
The last four rows and the effect of $G_3(0)$ are discussed
in the next sections.

The comparison between the first two rows
demonstrates how important the inclusion
of the experimental value of $A_{1/2} (0)$, is instead of $A_{1/2} (0) =0$,
in the determination of the first two elementary form factors.
The effect can also be seen
in the results for $G_M(0)$ and $G_E(0)$.
The differences are about 20\% for the magnetic form factor and 10\% for
the electric form factor.

In the next sections, we use the estimated values
for $G_1$ and $G_2$ at $Q^2=0$ to test if we can derive parametrizations
that may explain the experimental data for the amplitudes $A_{1/2}$ and $A_{3/2}$,
up to a certain range of $Q^2$.
Due to the approximated character of the parametrizations, 
we restrict the analysis to the region $Q^2 < 1$ GeV$^2$.
To estimate the uncertainties of the parametrizations,
we calculate also the uncertainties of $G_1$ and $G_2$ at $Q^2=0$,
based on the relations (\ref{eqG1a}) and (\ref{eqG2a})
and the data (\ref{eqAmpsEXP}),
with the errors combined in quadrature.
The numerical values for the uncertainties of $G_1(0)$
and $G_2(0)$  are included in the last row of Table~\ref{table-Models} 
(between brackets).
The relative errors are 7.8\% for $G_1$ and 10.6\% for $G_2$.

\section{Form factors $G_i$ for $Q^2 > 0$ \label{secResults2}}

In this section, we discuss possible parametrizations of the
amplitudes $A_{1/2}$ and $A_{3/2}$ for $Q^2 \le 1$ GeV$^2$
based on the values of $G_1(0)$ and $G_2(0)$
calculated in the previous section.

In the following, we consider the $\gamma^\ast N \to N(1520)$
helicity amplitude data from experiments
at JLab/CLAS on single-pion electroproduction~\cite{CLAS09}
and on charged double-pion electroproduction~\cite{CLAS12,CLAS16},
and the PDG selection for $Q^2=0$~\cite{PDG2022}.
These JLab/CLAS experiments determine the whole set of helicity amplitudes
($A_{1/2}$, $A_{3/2}$, and $S_{1/2}$).
The $\pi N$ ($\sim 60\%$) and the  $\pi \pi N$ ($\sim 30\%$) channels
are the dominant $N(1520)$ decay channels~\cite{PDG2022}.
There are additional data associated with different
experiments for the transverse amplitudes~\cite{Burkert03},
but the data analysis is based on the assumption
that $S_{1/2} \equiv 0$, an approximation
that is not valid at low $Q^2$~\cite{CLAS09}.

In a first stage, we ignore the role of the form factor $G_3$,
setting $G_3 \equiv 0$, 
since no information about $G_3$ can be obtained from the
transverse amplitudes at $Q^2=0$.
We notice, however, that $G_3$ contributes to the amplitudes
$A_{1/2}$ and $A_{3/2}$ for $Q^2 \ne 0$,
since $T_1$ and $T_2$ include the term $4 Z_R  Q^2 G_3 $
[see Eqs.(\ref{eqT1x}) and (\ref{eqT2x})].
Later on, we estimate the impact of nonzero values for $G_3(0)$.

From the previous section, we concluded already
that  $A_{1/2} (0) \simeq 0$ is not a very good approximation.
In the following, we consider then parametrizations
based on Eqs.~(\ref{eqG1a}) and (\ref{eqG2a})
consistent with the experimental value of $A_{1/2}(0)$. 
The numerical values are included
in the lower part of Table~\ref{table-Models} (with $T_1 = 0.319$).

We divided our analysis into several steps.

\subsection{Parametrization with constant form factors $G_i$}

The simplest parametrization can be obtained assuming
that the form factors $G_1(Q^2)$ and $G_2(Q^2)$ do not vary significantly
in the region $Q^2=0$...1 GeV$^2$
(meaning that in that range the derivatives of those
form factors are zero or negligible).
We label this approximation as the constant form factor parametrization.
The values of $G_i(0)$ are the ones presented
on Table~\ref{table-Models} in the first row with $T_1 = 0.319$.
As mentioned already, we assume for now that $G_3 (0)=0$.

\begin{figure*}[t]
%\vspace{.1cm}
\begin{center}
\mbox{
\includegraphics[width=3.1in]{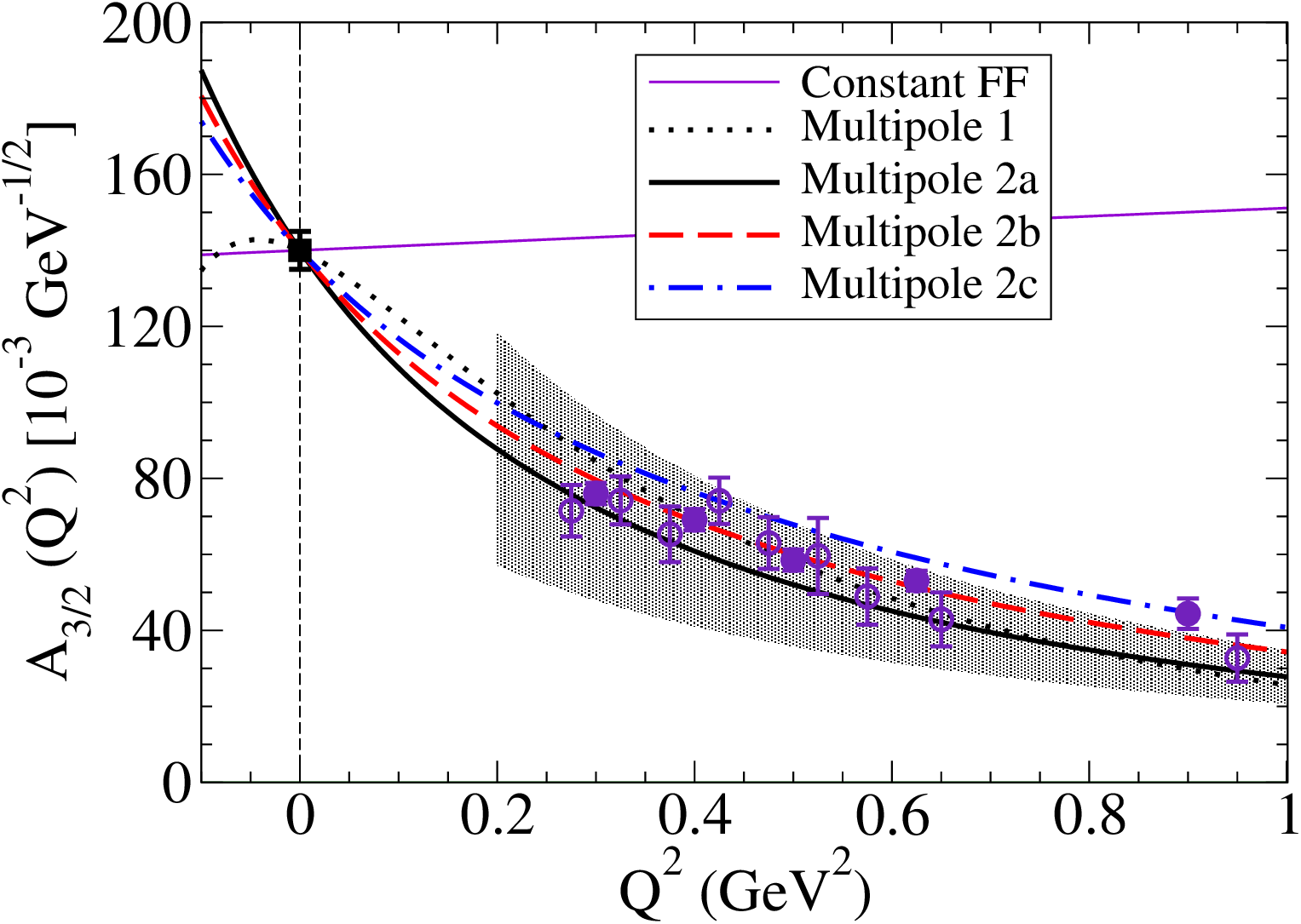} \hspace{.7cm}
\includegraphics[width=3.1in]{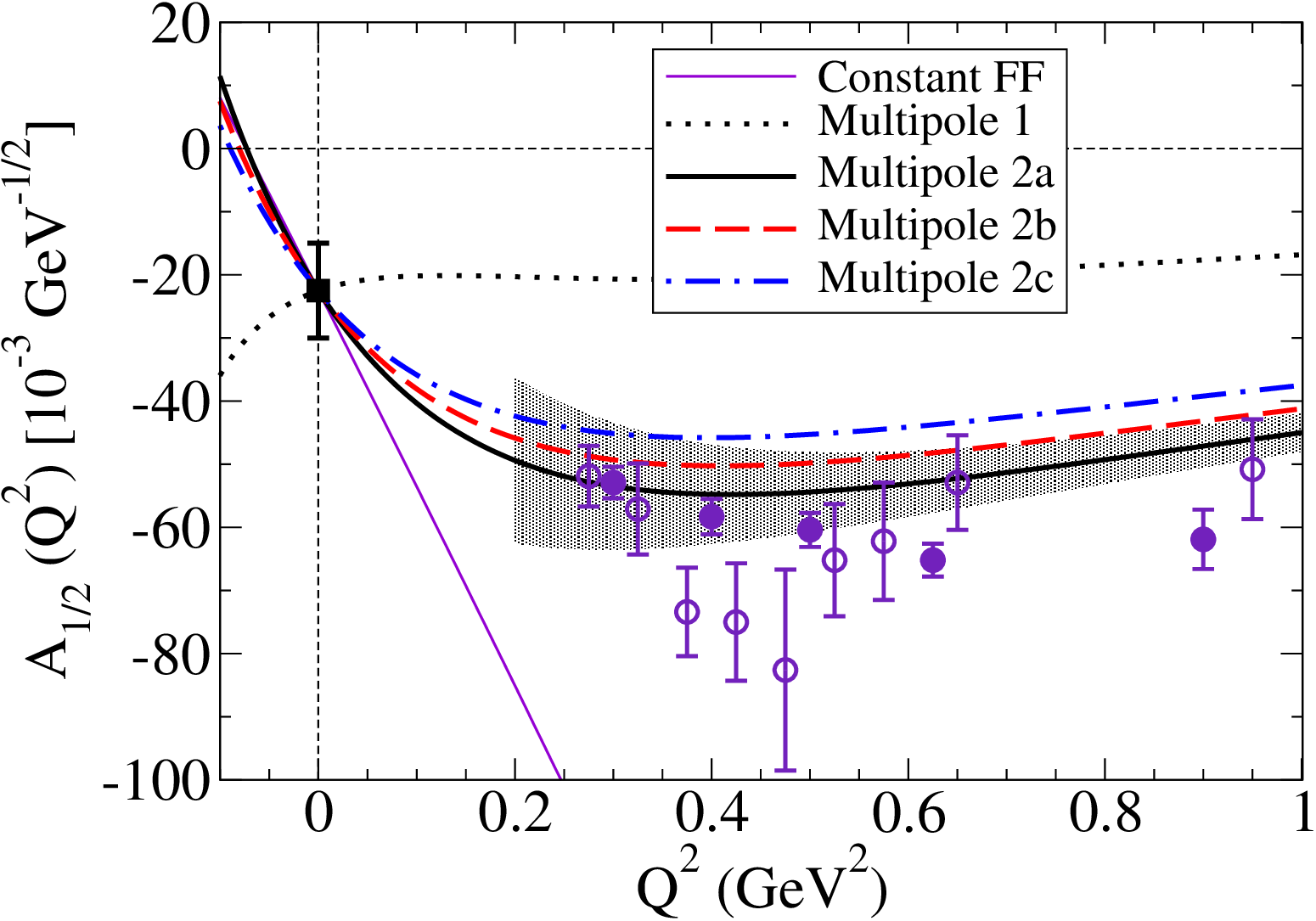} }
\end{center}
\caption{\footnotesize{Transverse amplitudes $A_{1/2}$ and $A_{3/2}$
    in terms of $Q^2$.
    The data are from JLab/CLAS single-pion production
    (solid bullets)~\cite{CLAS09},
    JLab/CLAS double-pion production (empty bullets)~\cite{CLAS12,CLAS16}
    and PDG (square)~\cite{PDG2022}.
    The labels correspond to the parameters from Table~\ref{table-Models}.
}}
\label{fig-Amps}
\end{figure*}

\begin{figure*}[t]
\vspace{.5cm}
\begin{center}
\mbox{
\includegraphics[width=3.1in]{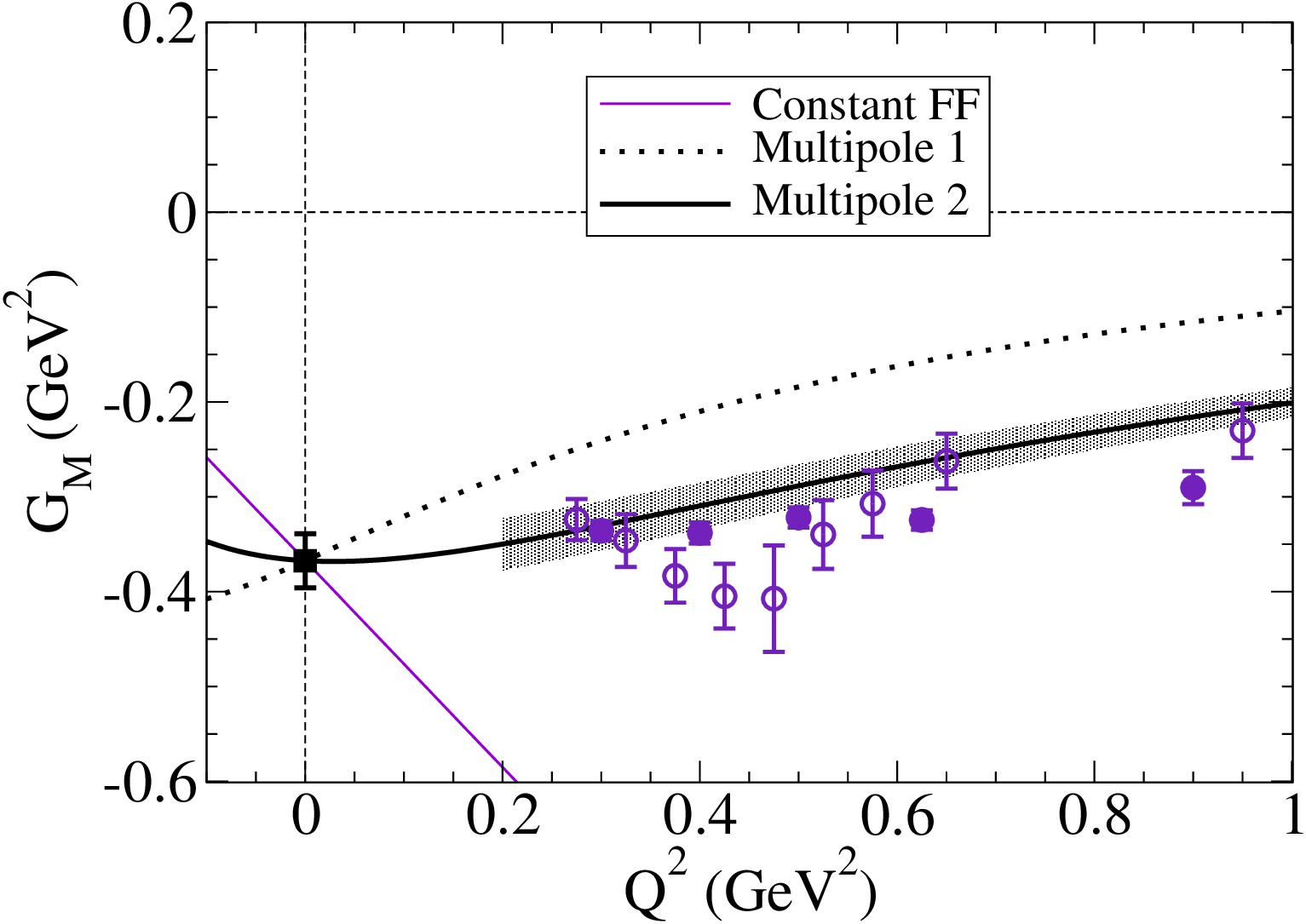} \hspace{.7cm}
\includegraphics[width=3.1in]{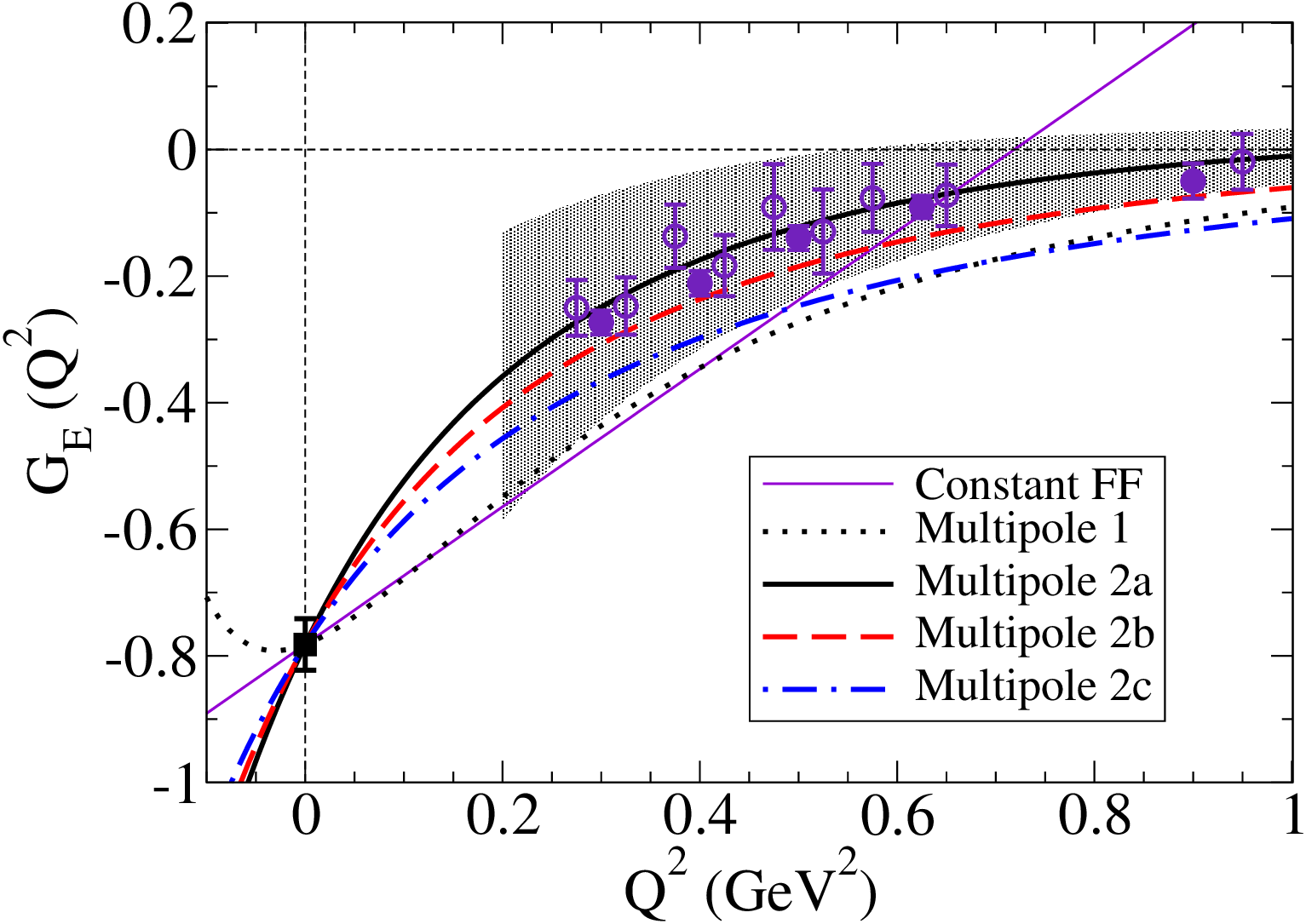} }
\end{center}
\caption{\footnotesize{Multipole form factors $G_E$ and $G_M$
    in terms of $Q^2$.
    Data description as in Fig.~\ref{fig-Amps}.
    The labels correspond to the parameters from Table~\ref{table-Models}.}}
\label{fig-FFs}
\end{figure*}

The constant form factor estimates 
are presented in Fig.~\ref{fig-Amps} for the amplitudes
and in Fig.~\ref{fig-FFs} for the multipole form factors.
Notice in the figures the lack of data for the interval $Q^2=0$...0.28 GeV$^2$.
This omission causes difficulty in the determination of the shape of
the helicity amplitudes near the photon point~\cite{PPNP2023,Siegert4}.

In Figs.~\ref{fig-Amps} and \ref{fig-FFs}, we distinguish between
the CLAS data from single-pion production~\cite{CLAS09}
from double-pion production~\cite{CLAS12,CLAS16}.
Some differences between the two sets can be
observed for the function $A_{3/2}$ and $G_M$
in the range $Q^2=0.35$...0.50 GeV$^2$.
The data are, however, compatible within the two standard deviation range.
More accurate data in that range may help
to determine the shape of the transverse amplitudes at low $Q^2$.

We present the calculations 
the range $Q^2=-0.1$...1 GeV$^2$ for a better visualization
of the results near $Q^2=0$.
The lower limit of the graph in $Q^2$
can be extended down to the pseudothreshold $Q^2 \simeq-0.4$ GeV$^2$,
in order to visualize the consequences
of the pseudothreshold constraints.
We notice, however, that the pseudothreshold conditions are
automatically satisfied by the use of 
elementary form factors $G_i$
when they have no singularities in the range $Q^2 > -(M_R-M)^2$.
At the end, we discuss the properties of $G_M$
near the pseudothreshold.

Before discussing the amplitudes $A_{1/2}$ and $A_{3/2}$
it is important to discuss the properties of the
form factors $G_E$ and $G_M$ when the form factors $G_i$ are constants.
From the relations (\ref{eqGE1}) and (\ref{eqGM1}), we can conclude
that the multipole form factors ($G_E$ and $G_M$)
are linear functions of $Q^2$.

As for the transverse amplitudes, we can notice that
they are  written 
in the form $A_{1/2} \propto \sqrt{1 + \tau} \;T_1$
and $A_{3/2} \propto \sqrt{1 + \tau} \; T_2$, where
$T_1$ and $T_2$ are linear functions and $\tau= Q^2/(M_R + M)^2$
[see Eqs.~(\ref{eqAmps-v1})--(\ref{eqFm})].
Since in the region of study $\tau \ll 1$
(because $Q^2 \ll 6.05$ GeV$^2$), 
one can write  $\sqrt{1 + \tau} \simeq 1 + \frac{1}{2} \frac{Q^2}{(M_R + M)^2}$,
and conclude that in the region  $Q^2 \le 1$ GeV$^2$, 
the amplitudes are well approximated by linear functions.

We can now discuss the numerical results for the amplitudes
$A_{1/2}$ and $A_{3/2}$ within the constant form factor approximation.
The estimates are presented in Fig.~\ref{fig-Amps} by the thin solid line
(labeled as Constant FF).
The model estimate $A_{3/2} \simeq 140$ $\times 10^{-3}$ GeV$^{-1/2}$,
contrasts with the sharp suppression of the experimental amplitude,
when $Q^2$ increases.
The estimate of  $A_{3/2}$ manifests only a weak dependence on $Q^2$, 
because $A_{3/2} \propto \sqrt{1 + \tau} \; T_2$ and $T_2$ is a constant.
The conclusion is then that the amplitude $A_{3/2}$ follows 
$A_{3/2}  \propto 1 + \frac{1}{2} \frac{Q^2}{(M_R + M)^2}$, in
the range of study, an almost constant function.
As for the amplitude $A_{1/2}$, we observe also an almost 
linear function\footnote{The term in $Q^4$
is very small because it is proportional to \mbox{$1/(M_R +M)^4$.}}
of $Q^2$, in flagrant disagreement with the data.
The conclusion is then, that the constant form factor approximation for 
the functions $G_i$ fails
completely the description of the amplitudes $A_{1/2}$ and $A_{3/2}$.

The corresponding results for $G_E$ and $G_M$
are presented in Fig.~\ref{fig-FFs}.
In this case, one obtains linear functions, which fail
in general the description of the data.

The corollary of this first analysis is that, the description
of the transverse amplitudes requires in addition
to the values of the functions $G_1$ and $G_2$ at $Q^2=0$,
the determination of the scale of variation of those functions.
In simple terms, we need an estimate of the
derivatives of the form factors $G_1$, $G_2$ and eventually $G_3$,
if we want to describe the data in the range $Q^2=0$...1 GeV$^2$.

\subsection{Parametrization of $G_i$ by multipole functions --
  universal cutoff}

Once concluded that the data are not consistent with
parametrizations based on constant elementary form factors,
we look for parametrizations based on multipole functions.
These kinds of parametrizations are considered, for instance
in the study of the nucleon electromagnetic 
form factors, where the main dependence is 
regulated by a simple dipole function.
Based on the expected asymptotic dependence of the functions $G_i$
in the large-$Q^2$ region, we consider the multipole parametrizations
\ba
G_1 (Q^2) = \frac{G_1(0)}{\left(
1 + \frac{Q^2}{\Lambda_3^2}
\right)^3},
\hspace{.6cm}
G_{2,3} (Q^2) = \frac{G_{2,3}(0)}{\left(
1 + \frac{Q^2}{\Lambda_4^2}
\right)^4},
\nonumber \\
\label{eqFFmultipoles1}
\ea
where we labeled the square form factor cutoffs $\Lambda_n^2$
by the power $n$ of the multipole.
We assume then that $G_2$ and $G_3$ (when $G_3(0) \ne 0$)
are regulated by the same cutoff.
The powers of Eq.~(\ref{eqFFmultipoles1}) are the ones compatible
with the expected falloffs for the helicity amplitudes (\ref{eqAmpsLQ2}) and
the multipole form factors (\ref{eqFFLQ2a}) and (\ref{eqFFLQ2b}).

The multipole functions take into account implicitly,
the leading-order dependence of the form factors $G_i$ on $Q^2$.
The method had been used in chiral effective-field theory
to include next-leading-order contributions
and improve the convergence of the calculations~\cite{Pascalutsa06a}.
It is also known that simple smooth 
parametrizations of $\gamma^\ast N \to N^\ast$ data are obtained
for most low-lying nucleon resonances
when the functions are normalized by an
appropriated multipole~\cite{Eichmann18a}.

One of the simplest parametrizations is obtained
when we assume that the scale of variation
of the form factors $G_i$ (associated with the square cutoffs
$\Lambda_3^2$ and $\Lambda_4^2$) can be the same
for all the form factors ($\Lambda_3^2= \Lambda_4^2$).
The condition $\Lambda_3^2= \Lambda_4^2$ defines the universal cutoff approximation.

Inspired by the nucleon dipole function
\ba
G_D (Q^2) = \frac{1}{\left(1 + \frac{Q^2}{\Lambda_D^2} \right)^2},
\label{eqDipole}
\ea
where $\Lambda_D^2 =0.71$ GeV$^2$,
we consider a parametrization where 
\ba
\Lambda_3^2 = \Lambda_4^2 = \Lambda_D^2.
\label{eqUniversal}
\ea

The results of the universal form factor parametrization
are represented in Figs.~\ref{fig-Amps} and \ref{fig-FFs} by the dotted line
and are labeled as Multipole 1.
We can notice in the figure for the amplitudes (Fig.~\ref{fig-Amps}), 
the failure in the description of the amplitude $A_{1/2}$.
Also worth noticing is the shape the amplitude $A_{3/2}$ near $Q^2=0$.
Although no data exist below $Q^2=0.28$ GeV$^2$, 
theoretical models predict in general a sharp
and fast falloff of the amplitude near $Q^2=0$.
In contrast, the line Multipole 1 has an almost zero derivative at $Q^2=0$.

In the constant-cutoff approximation,
we can also treat the cutoff $\Lambda_3 =\Lambda_4$
as an adjustable parameter, different from $\Lambda_D$,
to be determined by a fit to the data.
Different values of the cutoffs lead, however, to similar results.
The combination of the form factors $G_1$ and $G_2$
on $T_1$ and $T_2$ is such that it generates 
an almost constant estimate for $A_{1/2}$, and an almost zero derivative
for $A_{3/2}$ near $Q^2=0$.

The conclusion of this section is then that the data are not consistent with
multipole parametrizations based on the same cutoff for $G_1$ and $G_2$.

\subsection{Parametrization of $G_i$ by multipole functions --
  natural scale}

Since the universal cutoff fails to provide a
description of the low $Q^2$ transverse amplitude data,
we look for alternative ways of defining 
the scale of variation of the elementary form factors $G_i$.
Recalling that the nucleon elastic form factors 
and some inelastic transitions, such as
the $\gamma^\ast N \to \Delta (1232)$ magnetic form factor, 
scale at sufficient small $Q^2$ with the dipole function (\ref{eqDipole}),
we wondered if the same scale can be used for the functions $G_i$.
Since the functions are defined by different powers for the multipoles,
the similarity of the functions  $G_i$ with $G_D$ must
be imposed for low $Q^2$.
We consider then the conditions near $Q^2=0$,
\ba
\left(
1 + \frac{Q^2}{\Lambda_3^2}
\right)^{-3} &\simeq & 
\left(1 + \frac{Q^2}{\Lambda_D^2} \right)^{-2}, \\
\left(
1 + \frac{Q^2}{\Lambda_4^2}
\right)^{-4} &\simeq & 
\left(1 + \frac{Q^2}{\Lambda_D^2} \right)^{-2},
\ea
The equivalence of the previous expansions near $Q^2=0$ implies that
\ba
\Lambda_3^2 = \frac{3}{2} \Lambda_D^2,
\hspace{.6cm}
\Lambda_4^2 = 2\Lambda_D^2.
\label{eqMult2}
\ea
Numerically, one has $\Lambda_3^2 \simeq 1.07$ GeV$^2$ and
$\Lambda_4^2 \simeq 1.42$ GeV$^2$.

The numerical results associated with the multipole parametrization (\ref{eqFFmultipoles1})
with the cutoffs (\ref{eqMult2}) and $G_3(0)=0$ are presented in
Figs.~\ref{fig-Amps} and \ref{fig-FFs} by the thick solid line,
and are labeled as Multipole 2a.
Notice the closeness between the lines and the data.

Concerning the results from Fig.~\ref{fig-FFs} for $G_M$, a note is in order.
Since $G_M$ depend only on $G_1$,
all estimates discussed in this section have the same result for $G_M$
(thick solid line).
In the figure, we use the label Multipole 2.

The results of the parametrization Multipole 2a
demonstrate that a reliable description of
the $\gamma^\ast N \to N(1520)$ transverse amplitude data
can be achieved when we assume the natural scale for the
elementary form factors $G_i$.

We can now discuss the effect of the form factor $G_3$
in parametrizations of the data based on multipole functions.
Although $G_3(0)$ cannot be determined by the $A_{1/2}(0)$ and  $A_{3/2}(0)$ data,
indirect information can be obtained from the amplitude $S_{1/2}$ at low $Q^2$.
Unfortunately, no data below $Q^2=0.28$ GeV$^2$ are available to make a reliable
estimate of $S_{1/2} (0)$, and consequently an estimate of $G_3(0)$.

In these conditions, one has to rely on theoretical extrapolations of the data.
We consider then a parametrization of
the data from Ref.~\cite{Siegert4}, 
compatible with the pseudothreshold constraints
of the helicity amplitudes, and also with 
the low-$Q^2$ data for $A_{1/2}$ and $A_{3/2}$.
The value of $S_{1/2} (0)$ determined by that parametrization
is $S_{1/2} (0) = -64.4 \times 10^{-3}$ GeV$^{-1/2}$.
Combining this result with the present estimates
of $G_1(0)$ and $G_2(0)$, one obtains $M^2 G_3 (0)= -0.278$.

In addition to the parametrization discussed earlier
(Multipole 2a, $G_3(0)=0$), we consider also 
a parametrization with an intermediate value for $G_3(0)$,
fixed by $M^2 G_3 (0)= -0.14$, labeled as Multipole 2b,
and a parametrization associated with 
value of $S_{1/2} (0)$ mentioned above (Multipole 2c).
All parameters and associated values
for $S_{1/2} (0)$ are presented in
the last four rows of Table~\ref{table-Models}.

The parametrizations labeled as Multipole 2b and Multipole 2c
are also represented in Figs.~\ref{fig-Amps} and~\ref{fig-FFs}
by the dashed lines (Multipole 2b) and the dashed-dotted lines (Multipole 2c).

Since these parametrizations (Multipole 2a, 2b, 2c)
are defined by the values of $G_1(0)$ and $G_2(0)$
determined by the transverse amplitudes at $Q^2=0$, 
one can also calculate the band of variation of
the parametrizations based on the uncertainties of the parameters.
For clarity, we include only the band of variation
associated with the Multipole 2a.
The others have similar ranges of variation from the central lines.
The bands of variation are large for estimates near $Q^2=0$, 
when the errors are added in quadrature,
mainly due to the large relative uncertainty of $A_{1/2} (0)$.
For that reason, we restrict the representation to $Q^2 \ge 0.2$ GeV$^2$.
The width of the bands decreases when $Q^2$ increases due to 
the reduction of the values of the functions $G_i$.
More accurate experimental estimates of $A_{1/2} (0)$ and $A_{3/2} (0)$
will narrow the uncertainties of the estimates based
on Eqs.~(\ref{eqG1a}),  (\ref{eqG2a}), 
(\ref{eqFFmultipoles1}) and (\ref{eqMult2}).

From the analysis of the amplitudes (Fig.~\ref{fig-Amps}), we
can conclude that the best description of the amplitude $A_{1/2}$
is obtained with Multipole 2a ($G_3=0$).
Notice, however, that Multipole 2b provides also
a fair description of the data when the uncertainties are taken into account.
As for the amplitude $A_{3/2}$, Multipole 2b gives
the best description when we consider the central values,
but Multipole 2a and Multipole 2c are also consistent with the data,
when we take into account the uncertainties
(upper error band for Multipole 2a and lower error band for  Multipole 2c).
Overall Multipole 2a and Multipole 2b give the best combined
description of the transverse amplitudes within the uncertainty bands.
The agreement with the data is better for $Q^2 \le 0.7$ GeV$^2$.

The preference for the parametrizations Multipole 2a and Multipole 2b,
favors also models with large magnitudes for the absolute values
of the scalar amplitude $S_{1/2}(0)$,
associated with the range $-$(75...85)$\times 10^{-3}$ GeV$^{-1/2}$,
as indicated in Table~\ref{table-Models}.

Similar conclusions are obtained when we look
for the multipole form factors $G_E$ and $G_M$ (Fig.~\ref{fig-FFs}).
All parametrizations are equivalent for $G_M$.
The data for $G_E$ favor the parametrizations
Multipole 2a and Multipole 2b, within the intervals of variation.

In the graph for $G_M$,
one can also observe that the function is very smooth near $Q^2=0$,
contrasting with the sharp variation of $G_E$.
This effect is a consequence of the particular
condition for $G_M$ at the pseudothreshold,
as discussed in Sec.~\ref{secFormalismA}.
No equivalent condition exists for $G_E$ and $G_C$
(related by $G_E \propto G_C$).
Both functions, $G_E$ and $G_C$, are finite at the pseudothreshold.

A consequence of the condition $G_M=0$ at the pseudothreshold
is that we can expect a turning point of the function
below $Q^2=0.2$ GeV$^2$.
The present calculations suggest that the turning point
is close to $Q^2=0$, meaning that the derivative of $G_M$
at photon point is close to zero. 
Considering the relation between $G_M$ and $G_1$,
we can conclude that $(M_R-M)^2 \frac{d G_M}{d Q^2} (0) =
\left(1 - 3\frac{(M_R-M)^2}{\Lambda_3^2} \right) G_M (0) \simeq 0.05 G_M(0)$,
consistent with a very small value for $\frac{d G_M}{d Q^2}(0)$.
This result is a direct consequence of
the parameter $\Lambda_3^2 = 1.07$ GeV$^2$.

The parametrizations discussed above can also be compared with
recent parametrizations proposed in the literature.
Of particular interest is the parametrization
from Refs.~\cite{HBlin19a,Jlab-fits}
mentioned here as the JLab parametrization.
The JLab parametrization is based on rational functions
calibrated by JLab/CLAS and PDG ($Q^2=0$) data.
The parametrization is close to the Multipole 2a parametrization
within one standard deviation for $A_{3/2}$ and one
or two standard deviations for $A_{1/2}$ 
in the $Q^2 < 0.8$ GeV$^2$ region (estimated by Multipole 2a).
It provides also a good description of the large-$Q^2$ data.

The JLab parametrization has an important property:
although the extension of the parametrization
to the $Q^2 < 0$ region is not compatible with
the pseudothreshold constraints, it can be analytically continued
to the timelike region, in order to fulfill
the pseudothreshold constraints~\cite{Siegert4}.
This analytic continuation provides a soft transition
between the region $Q^2 \simeq 0$ and the
pseudothreshold and leaves the original parametrization of the
region $0 \le Q^2 \le 0.8$ GeV$^2$ almost unchanged.
Overall, one obtains a parametrization consistent with
the low-$Q^2$ data and the necessary pseudothreshold constraints, 
preserving at the same time the original form
for the large-$Q^2$ region~\cite{Siegert4}.

%% XVSPACE

%\vspace{-.6cm}
\subsection{Discussion}
%\vspace{-.2cm}

The parametrizations discussed above are based 
on two parameters, $G_1(0)$ and $G_2(0)$,
a cutoff determined by theoretical arguments,
and some tentative estimates of $G_3(0)$.
Two of the parametrizations provide
good descriptions of the data for $A_{1/2}$ and $A_{3/2}$ for $Q^2 < 1$ GeV$^2$,
and determine also the possible range variation 
for the amplitude $S_{1/2}$ near $Q^2=0$.

From our analysis, we conclude also 
that the data favor parametrizations with multipole functions
regulated by large cutoffs ($\Lambda_3^2$, $\Lambda_4^2 > 1$ GeV$^2$)
and slower falloffs.
The considered cutoffs are larger
than the cutoff associated with the nucleon
elastic form factors ($\Lambda_D^2 \simeq 0.7$ GeV$^2$). 

In principle, more accurate estimates can be obtained
considering extensions of the multipole parametrizations,
where the second derivatives of $G_i$ are adjusted by the low-$Q^2$ data.
We did not test this possibility, because
the main goal of the present work is the understanding of
the $Q^2=0$ and low-$Q^2$ data based on a minimal number
of parameters and assumptions.

The parametrizations proposed here may be tested in the near future
by experiments in the range $Q^2=0$...0.3 GeV$^2$,
in order to fill the gap in the experimental studies
of the $N(1520)$ resonance.
Those data may be acquired at MAMI ($Q^2> 0.2$ GeV$^2$)
and JLab ($Q^2 > 0.05$ GeV$^2$)~\cite{Mokeev22a,CLAS23a}.

New data can help to determine the shape
of the transverse amplitudes below $Q^2=0.3$ GeV$^2$,
and impose more accurate constraints on
parametrizations of the data near $Q^2=0$~\cite{Siegert4}.
The knowledge of the $Q^2$ dependence of the helicity amplitudes
near $Q^2=0$ is important for the study
of the $\gamma^\ast N \to N(1520)$ in the timelike region ($Q^2 < 0$),
including the Dalitz decay of the $N(1520)$ state
[$N(1520) \to e^+ e^- N$]~\cite{N1520,HADES}.

%\vspace{-.5cm}
\section{Outlook and conclusions \label{secConclusions}}

The $N(1520)$ resonance is among the
nucleon excitations that are better known experimentally.
It differs from the other low-lying nucleon resonances
by its properties.
The transverse amplitudes $A_{1/2}$ and $A_{3/2}$ have completely
different magnitudes at $Q^2=0$, and are subject
to relevant constraints at low $Q^2$, 
due to the proximity between the pseudothreshold
$Q^2=-(M_R-M)^2$ and the photon point.

In the present work, we looked for the origin
of the difference of magnitudes between
the transverse amplitudes at very low $Q^2$.
We concluded that the result is related to a significant
cancellation near $Q^2=0$ of the contributions associated with two
elementary form factors ($G_1$ and $G_2$), defined by a gauge-invariant
parametrization of the transition current. 
We concluded also that, the correlation between
the elementary form factors does not hold for larger values of $Q^2$.

To explain the shape of the amplitudes $A_{1/2}$ and $A_{3/2}$
below $Q^2=1$ GeV$^2$, in addition to the values of $G_1$ and $G_2$ at $Q^2=0$,
one needs to know the scale of variation of
the elementary form factors $G_1$, $G_2$, and $G_3$.
We obtain a fair description of the $Q^2 \le 0.7$ GeV$^2$ data
when the scale of variation of the elementary form factors
is correlated to the natural scale of
the $\gamma^\ast N \to N^\ast$ transition amplitudes,
defined by the nucleon dipole form factor.

Different parametrizations can be derived
depending on the projected value of
the scalar amplitude $S_{1/2}$ near $Q^2=0$.
Those parametrizations are compatible
with the experimental data for the transverse amplitudes
within the uncertainties of the data for $Q^2=0$.
The uncertainties can be reduced
once the more accurate determinations of
the transverse amplitudes are provided, mainly for $A_{1/2} (0)$.
The proposed parametrizations explain also the smooth behavior of
the magnetic dipole form factor $G_M$ near $Q^2=0$, suggested by the data.

Our analysis of the transverse amplitudes $A_{1/2}$ and $A_{3/2}$
for finite $Q^2$ allows us to make an estimate of 
the range of variation of the scalar amplitude  $S_{1/2}$ near $Q^2=0$,
in a region for which there are no data available.
Our parametrizations are compatible
with values of $S_{1/2} (Q^2)$ in the range from $-85 \times 10^{-3}$ GeV$^{-1/2}$
to $-75 \times 10^{-3}$ GeV$^{-1/2}$, for values of 
$Q^2$ near the photon point.  
The parametrizations discussed in the present work may be tested in
future measurements of the transverse and
longitudinal amplitudes for $0 < Q^2 < 0.28$ GeV$^2$.

\vspace{1.cm}

\begin{acknowledgments}
 %\vspace{-.4cm}
G.~R.~was supported the Basic Science Research Program 
through the National Research Foundation of Korea (NRF)
funded by the Ministry of Education  (Grant No.~NRF–2021R1A6A1A03043957).
\end{acknowledgments}


\begin{thebibliography}{00}


  
\bibitem{PPNP2023}
G.~Ramalho and M.~T.~Pe\~na,
%``Electromagnetic transition form factors of baryon resonances,''
Prog. Part. Nucl. Phys. \textbf{136}, 104097 (2024)
%%doi:10.1016/j.ppnp.2024.104097
[arXiv:2306.13900 [hep-ph]].
%8 citations counted in INSPIRE as of 15 Mar 2024
  

\bibitem{Aznauryan12} 
  I.~G.~Aznauryan and V.~D.~Burkert,
  %``Electroexcitation of nucleon resonances,''
  Prog.\ Part.\ Nucl.\ Phys.\  {\bf 67}, 1 (2012)
  %% doi:10.1016/j.ppnp.2011.08.001
  [arXiv:1109.1720 [hep-ph]].
  %%CITATION = doi:10.1016/j.ppnp.2011.08.001;%%
  %163 citations counted in INSPIRE as of 10 Jan 2020



\bibitem{NSTAR} 
  I.~G.~Aznauryan {\it et al.},
  %``Studies of Nucleon Resonance Structure in Exclusive Meson Electroproduction,''
  Int.\ J.\ Mod.\ Phys.\ E {\bf 22}, 1330015 (2013)
  %% doi:10.1142/S0218301313300154
  [arXiv:1212.4891 [nucl-th]].
  %%CITATION = doi:10.1142/S0218301313300154;%%
  %63 citations counted in INSPIRE as of 18 Dec 2015


  
\bibitem{CLAS23a}
V.~I.~Mokeev, P.~Achenbach, V.~D.~Burkert, D.~S.~Carman, R.~W.~Gothe, A.~N.~Hiller Blin, E.~L.~Isupov, K.~Joo, K.~Neupane and A.~Trivedi,
%``First results on nucleon resonance electroexcitation amplitudes from ep\textrightarrow{}e'\ensuremath{\pi}+\ensuremath{\pi}\ensuremath{-}p' cross sections~at W=1.4\textendash{}1.7 GeV and Q2=2.0\textendash{}5.0GeV2,''
Phys. Rev. C \textbf{108}, 025204 (2023)
%%doi:10.1103/PhysRevC.108.025204
[arXiv:2306.13777 [nucl-ex]].
%1 citations counted in INSPIRE as of 07 Oct 2023


\bibitem{Drechsel07a}
D.~Drechsel, S.~S.~Kamalov and L.~Tiator,
%``Unitary Isobar Model - MAID2007,''
Eur. Phys. J. A \textbf{34}, 69 (2007)
%%doi:10.1140/epja/i2007-10490-6
[arXiv:0710.0306 [nucl-th]].
%551 citations counted in INSPIRE as of 21 Oct 2023


\bibitem{Burkert04} 
  V.~D.~Burkert and T.~S.~H.~Lee,
  %``Electromagnetic meson production in the nucleon resonance region,''
  Int.\ J.\ Mod.\ Phys.\ E {\bf 13}, 1035 (2004)
  %% doi:10.1142/S0218301304002545
  [nucl-ex/0407020].
  %%CITATION = doi:10.1142/S0218301304002545;%%
  %186 citations counted in INSPIRE as of 14 Jan 2020


\bibitem{Siegert4} 
  G.~Ramalho,
  %``Low-$Q^2$ empirical parametrizations of the $N^\ast$ helicity amplitudes,''
  Phys.\ Rev.\ D {\bf 100}, 114014 (2019)
  %%doi:10.1103/PhysRevD.100.114014
  [arXiv:1909.00013 [hep-ph]].
  %%CITATION = doi:10.1103/PhysRevD.100.114014;%%
  %11 citations counted in INSPIRE as of 10 Jan 2020



\bibitem{Tiator16a}
L.~Tiator,
%``Pion Electroproduction and Siegert's Theorem,''
Few Body Syst. \textbf{57}, 1087 (2016).
%%doi:10.1007/s00601-016-1158-1
%13 citations counted in INSPIRE as of 07 Oct 2023



\bibitem{PDG2022}
R.~L.~Workman \textit{et al.} [Particle Data Group],
%``Review of Particle Physics,''
PTEP \textbf{2022}, 083C01 (2022).
%%doi:10.1093/ptep/ptac097
%138 citations counted in INSPIRE as of 18 Oct 2022



\bibitem{Mokeev22a}
V.~I.~Mokeev \textit{et al.} [CLAS],
%``Photo- and Electrocouplings of Nucleon Resonances,''
Few Body Syst. \textbf{63}, 59 (2022)
%%doi:10.1007/s00601-022-01760-2
[arXiv:2202.04180 [nucl-ex]].
%19 citations counted in INSPIRE as of 14 Oct 2023
%%  Limits of measurements



  
\bibitem{Devenish76} 
  R.~C.~E.~Devenish, T.~S.~Eisenschitz and J.~G.~Korner,
  %``Electromagnetic $N−N^∗$ transition form factors,''
  Phys.\ Rev.\ D {\bf 14}, 3063 (1976).
  %% doi:10.1103/PhysRevD.14.3063
  %%CITATION = doi:10.1103/PhysRevD.14.3063;%%
  %95 citations counted in INSPIRE as of 04 Dec 2015



\bibitem{Siegert3}
G.~Ramalho,
%``Improved empirical parametrizations of the $\gamma^\ast N \to \Delta(1232)$ and $\gamma^\ast N \to N(1520)$ helicity amplitudes and the Siegert's theorem,''
Phys. Rev. D \textbf{93}, 113012 (2016)
%%doi:10.1103/PhysRevD.93.113012
[arXiv:1602.03832 [hep-ph]].
%22 citations counted in INSPIRE as of 24 Sep 2023


\bibitem{Siegert2} 
  G.~Ramalho,
  %``Parametrizations of the $\gamma^\ast N \to \Delta(1232)$ quadrupole form factors and Siegert’s theorem,''
  Phys.\ Rev.\ D {\bf 94}, 114001 (2016)
  %%doi:10.1103/PhysRevD.94.114001
  [arXiv:1606.03042 [hep-ph]].
  %%CITATION = doi:10.1103/PhysRevD.94.114001;%%
  %21 citations counted in INSPIRE as of 24 Dec 2019


\bibitem{Siegert5}
G.~Ramalho,
%``Improved empirical parametrizations of the $\gamma^\ast N \to N(1535)$ transition amplitudes and the Siegert's theorem,''
Phys. Lett. B \textbf{759}, 126 (2016)
%%doi:10.1016/j.physletb.2016.05.060
[arXiv:1602.03444 [hep-ph]].
%19 citations counted in INSPIRE as of 10 Oct 2023
  


\bibitem{Siegert-note}
Siegert's theorem is a general condition valid for the
$\gamma^\ast N \to N^\ast$ transitions 
that relate the electric amplitude (combination of $A_{1/2}$ and $A_{3/2}$)
and the scalar/longitudinal amplitude $S_{1/2}$
near \mbox{$Q^2=-(M_R-M)^2$.}
  


\bibitem{Drechsel92a}
D.~Drechsel and L.~Tiator,
%``Threshold pion photoproduction on nucleons,''
J. Phys. G \textbf{18}, 449 (1992).
%doi:10.1088/0954-3899/18/3/004
%290 citations counted in INSPIRE as of 07 Oct 2023

  
\bibitem{Buchmann98a}
A.~J.~Buchmann, E.~Hernandez, U.~Meyer and A.~Faessler 
%``N --\ensuremath{>} Delta (1232) E2 transition and Siegert's theorem,''
Phys. Rev. C \textbf{58}, 2478 (1998).
%%doi:10.1103/PhysRevC.58.2478
%45 citations counted in INSPIRE as of 07 Oct 2023




\bibitem{Warns90a}
M.~Warns, W.~Pfeil and H.~Rollnik,
%``Helicity and Isospin Asymmetries in the Electroproduction of Nucleon Resonances,''
Phys. Rev. D \textbf{42}, 2215 (1990).
%%doi:10.1103/PhysRevD.42.2215
%66 citations counted in INSPIRE as of 25 Oct 2023


\bibitem{N1520}
  %\bibitem{N1520SL}   
  G.~Ramalho and M.~T.~Pe\~na,
  %``$\gamma^\ast N \to N^\ast(1520)$ form factors in the spacelike region,''
  Phys.\ Rev.\ D {\bf 89}, 094016 (2014)
  %% doi:10.1103/PhysRevD.89.094016
  [arXiv:1309.0730 [hep-ph]];
  %%CITATION = doi:10.1103/PhysRevD.89.094016;%%
  %29 citations counted in INSPIRE as of 03 May 2021
  %\bibitem{N1520TL} 
  G.~Ramalho and M.~T.~Pe\~na,
  %``γ*N→N*(1520) form factors in the timelike regime,''
  Phys.\ Rev.\ D {\bf 95}, 014003 (2017)
  %%doi:10.1103/PhysRevD.95.014003
  [arXiv:1610.08788 [nucl-th]].
  %%CITATION = doi:10.1103/PhysRevD.95.014003;%%
  %6 citations counted in INSPIRE as of 05 Dec 2019




\bibitem{CLAS16}
V.~I.~Mokeev, V.~D.~Burkert, D.~S.~Carman, L.~Elouadrhiri, G.~V.~Fedotov, E.~N.~Golovatch, R.~W.~Gothe, K.~Hicks, B.~S.~Ishkhanov, E.~L.~Isupov, \textit{et al.}
%``New Results from the Studies of the $N(1440)1/2^+$, $N(1520)3/2^-$, and $\Delta(1620)1/2^-$ Resonances in Exclusive $ep \to e'p' \pi^+ \pi^-$ Electroproduction with the CLAS Detector,''
Phys. Rev. C \textbf{93}, 025206 (2016)
%%doi:10.1103/PhysRevC.93.025206
[arXiv:1509.05460 [nucl-ex]].
%92 citations counted in INSPIRE as of 19 Oct 2023






\bibitem{Merten02}
D.~Merten, R.~Ricken, M.~Koll, B.~Metsch and H.~Petry,
%``Weak decays of heavy mesons in a covariant quark model,''
Eur. Phys. J. A \textbf{13}, 477 (2002)
%%doi:10.1007/s10050-002-8778-1
[arXiv:hep-ph/0104029 [hep-ph]].
%16 citations counted in INSPIRE as of 25 Oct 2023


\bibitem{Ronniger13a}
M.~Ronniger and B.~C.~Metsch,
%``Effects of a spin-flavour dependent interaction on light-flavoured baryon helicity amplitudes,''
Eur. Phys. J. A \textbf{49}, 8 (2013)
%%doi:10.1140/epja/i2013-13008-9
[arXiv:1207.2640 [hep-ph]].
%22 citations counted in INSPIRE as of 25 Oct 2023


\bibitem{Giannini15a}
M.~M.~Giannini and E.~Santopinto,
%``The hypercentral Constituent Quark Model and its application to baryon properties,''
Chin. J. Phys. \textbf{53}, 020301 (2015)
%%doi:10.6122/CJP.20150120
[arXiv:1501.03722 [nucl-th]].
%119 citations counted in INSPIRE as of 25 Oct 2023





\bibitem{Golli13a}
B.~Golli and S.~\v{S}irca,
%``A chiral quark model for meson electroproduction in the region of D-wave resonances,''
Eur. Phys. J. A \textbf{49}, 111 (2013)
%%doi:10.1140/epja/i2013-13111-y
[arXiv:1306.3330 [nucl-th]].
%20 citations counted in INSPIRE as of 25 Oct 2023



\bibitem{Bijker09a}
R.~Bijker and E.~Santopinto,
%``Unquenched quark model for baryons: Magnetic moments, spins and orbital angular momenta,''
Phys. Rev. C \textbf{80}, 065210 (2009)
%%doi:10.1103/PhysRevC.80.065210
[arXiv:0912.4494 [nucl-th]].
%94 citations counted in INSPIRE as of 25 Oct 2023




\bibitem{JDiaz08}
B.~Julia-Diaz, T.~S.~H.~Lee, A.~Matsuyama, T.~Sato and L.~C.~Smith,
%``Dynamical coupled-channels effects on pion photoproduction,''
Phys. Rev. C \textbf{77}, 045205 (2008)
%%doi:10.1103/PhysRevC.77.045205
[arXiv:0712.2283 [nucl-th]].
%103 citations counted in INSPIRE as of 25 Oct 2023

\bibitem{Kamano13}
H.~Kamano, S.~X.~Nakamura, T.~S.~H.~Lee and T.~Sato,
%``Nucleon resonances within a dynamical coupled-channels model of $\pi N$ and $\gamma N$ reactions,''
Phys. Rev. C \textbf{88}, 035209 (2013)
%%doi:10.1103/PhysRevC.88.035209
[arXiv:1305.4351 [nucl-th]].
%240 citations counted in INSPIRE as of 25 Oct 2023



\bibitem{SemiRel}  
  G.~Ramalho,
  %``Semirelativistic approximation to the $\gamma^\ast N \to N(1520)$ and $\gamma^\ast N \to N(1535)$ transition form factors,''
  Phys.\ Rev.\ D {\bf 95}, 054008 (2017)
  %%doi:10.1103/PhysRevD.95.054008
  [arXiv:1612.09555 [hep-ph]].
  %%CITATION = doi:10.1103/PhysRevD.95.054008;%%
  %20 citations counted in INSPIRE as of 03 May 2021


\bibitem{Jones73} 
  H.~F.~Jones and M.~D.~Scadron,
  %``Multipole gamma N Delta form-factors and resonant photoproduction and electroproduction,''
  Annals Phys.\  {\bf 81}, 1 (1973).
  %% doi:10.1016/0003-4916(73)90476-4
  %%CITATION = doi:10.1016/0003-4916(73)90476-4;%%
  %233 citations counted in INSPIRE as of 17 Dec 2015








% ----------------------------------------------------
  
\bibitem{DeForest66a}
T.~De Forest, Jr. and J.~D.~Walecka,
%``Electron scattering and nuclear structure,''
Adv. Phys. \textbf{15}, 1 (1966).
%%doi:10.1080/00018736600101254
%748 citations counted in INSPIRE as of 07 Oct 2023


  
\bibitem{Amaldi79a}
E.~Amaldi, S.~Fubini and G.~Furlan,
%``PION ELECTROPRODUCTION. ELECTROPRODUCTION AT LOW-ENERGY AND HADRON FORM-FACTORS,''
Springer Tracts Mod. Phys. \textbf{83}, 1 (1979).
%%doi:10.1007/BFb0048209
%49 citations counted in INSPIRE as of 07 Oct 2023

%-------------------------------------------------------------

  
\bibitem{Siegert1} 
  G.~Ramalho,
  %``New low-$Q^2$ measurements of the $\gamma^\ast N \to \Delta(1232)$ Coulomb quadrupole form factor, pion cloud parametrizations and Siegert's theorem,''
  Eur.\ Phys.\ J.\ A {\bf 54}, 75 (2018)
  %%doi:10.1140/epja/i2018-12514-6
  [arXiv:1709.07412 [hep-ph]].
  %%CITATION = doi:10.1140/epja/i2018-12514-6;%%
  %14 citations counted in INSPIRE as of 24 Dec 2019



  
  
\bibitem{Carlson8X}
    C.~E.~Carlson,
  %``Electromagnetic N - Delta Transition At High Q**2,''
  Phys.\ Rev.\ D {\bf 34}, 2704 (1986);
  %%doi:10.1103/PhysRevD.34.2704
  %%CITATION = doi:10.1103/PhysRevD.34.2704;%%
  %171 citations counted in INSPIRE as of 18 Mar 2020
  %\bibitem{Carlson88a}
C.~E.~Carlson and J.~L.~Poor,
%``Distribution Amplitudes and Electroproduction of the $\Delta$ and Other Low Lying Resonances,''
Phys. Rev. D \textbf{38}, 2758 (1988).
%%doi:10.1103/PhysRevD.38.2758
%97 citations counted in INSPIRE as of 09 Oct 2023


\bibitem{Carlson98b}
C.~E.~Carlson and N.~C.~Mukhopadhyay,
%``Bloom-Gilman duality in the resonance spin structure functions,''
Phys. Rev. D \textbf{58}, 094029 (1998)
%%doi:10.1103/PhysRevD.58.094029
[arXiv:hep-ph/9801205 [hep-ph]].
%60 citations counted in INSPIRE as of 09 Oct 2023


\bibitem{CLAS09}
I.~G.~Aznauryan \textit{et al.} [CLAS],
%``Electroexcitation of nucleon resonances from CLAS data on single pion electroproduction,''
Phys. Rev. C \textbf{80}, 055203 (2009)
%%doi:10.1103/PhysRevC.80.055203
[arXiv:0909.2349 [nucl-ex]].
%284 citations counted in INSPIRE as of 19 Oct 2023


\bibitem{CLAS12}
V.~I.~Mokeev \textit{et al.} [CLAS],
%``Experimental Study of the $P_{11}(1440)$ and $D_{13}(1520)$ resonances from CLAS data on $ep \rightarrow e'\pi^{+} \pi^{-} p'$,''
Phys. Rev. C \textbf{86}, 035203 (2012)
%%doi:10.1103/PhysRevC.86.035203
[arXiv:1205.3948 [nucl-ex]].
%132 citations counted in INSPIRE as of 19 Oct 2023




\bibitem{Burkert03}
V.~D.~Burkert, R.~De Vita, M.~Battaglieri, M.~Ripani and V.~Mokeev,
%``Single quark transition model analysis of electromagnetic nucleon resonance transitions in the [70,1-] supermultiplet,''
Phys. Rev. C \textbf{67}, 035204 (2003)
%%doi:10.1103/PhysRevC.67.035204
[arXiv:hep-ph/0212108 [hep-ph]].
%67 citations counted in INSPIRE as of 25 Oct 2023



\bibitem{Pascalutsa06a}
V.~Pascalutsa and M.~Vanderhaeghen,
%``Chiral effective-field theory in the Delta(1232) region: I. Pion electroproduction on the nucleon,''
Phys. Rev. D \textbf{73}, 034003 (2006)
%%doi:10.1103/PhysRevD.73.034003
[arXiv:hep-ph/0512244 [hep-ph]].
%114 citations counted in INSPIRE as of 11 Oct 2023


\bibitem{Eichmann18a}
G.~Eichmann and G.~Ramalho,
%``Nucleon resonances in Compton scattering,''
Phys. Rev. D \textbf{98}, 093007 (2018)
%%doi:10.1103/PhysRevD.98.093007
[arXiv:1806.04579 [hep-ph]].
%21 citations counted in INSPIRE as of 11 Oct 2023



\bibitem{HBlin19a}
A.~N.~Hiller Blin, V.~Mokeev, M.~Albaladejo, C.~Fern\'andez-Ram\'\i{}rez, V.~Mathieu, A.~Pilloni, A.~Szczepaniak, V.~D.~Burkert, V.~V.~Chesnokov, A.~A.~Golubenko, \textit{et al.}
%``Nucleon resonance contributions to unpolarised inclusive electron scattering,''
Phys. Rev. C \textbf{100}, 035201 (2019)
%%doi:10.1103/PhysRevC.100.035201
[arXiv:1904.08016 [hep-ph]].
%32 citations counted in INSPIRE as of 15 Mar 2024



\bibitem{Jlab-fits}
\url{https://userweb.jlab.org/~isupov/couplings/}.


\bibitem{HADES}
R.~Abou Yassine \textit{et al.} [HADES],
%``First measurement of massive virtual photon emission from N* baryon resonances,''
[arXiv:2205.15914 [nucl-ex]];
%8 citations counted in INSPIRE as of 01 Dec 2023
R.~Abou Yassine \textit{et al.} [HADES],
%``Inclusive e$^+$e$^-$ production in collisions of pions with protons and nuclei in the second resonance region of baryons,''
[arXiv:2309.13357 [nucl-ex]].
%0 citations counted in INSPIRE as of 01 Dec 2023



\end{thebibliography}
\end{document}